\documentclass[fleqn,usenatbib]{mnras}

\usepackage{newtxtext,newtxmath}
\usepackage[T1]{fontenc}

\DeclareRobustCommand{\VAN}[3]{#2}
\let\VANthebibliography\thebibliography
\def\thebibliography{\DeclareRobustCommand{\VAN}[3]{##3}\VANthebibliography}

\usepackage{graphicx}
\usepackage{amsmath}

\usepackage{amssymb}
\usepackage{soul}

\hypersetup{linkcolor=red,citecolor=red,filecolor=cyan,urlcolor=magenta}

\newcommand{\pcm}{\,pc\,cm$^{-3}$}	

\title[Ultrawideband observations of PSR J1401$-$6357]{Individual
pulse emission from the diffuse drifter PSR J1401$-$6357 using the ultrawideband 
receiver on the Parkes radio telescope}

\author[J. L. Chen et al.]{
J. L. Chen,$^{1}$
Z. G. Wen,$^{2,3,4,5}$\thanks{E-mail: wenzhigang@xao.ac.cn}
X. F. Duan,$^{2,4}$
D. L. He,$^{2}$
N. Wang,$^{2,3}$
H. G. Wang,$^{6,2}$
R. Yuen,$^{2}$
\and
J. P. Yuan,$^{2,3}$ 
W. M. Yan,$^{2,3}$
Z. Wang,$^{7,2}$
C. B. Lv,$^{6}$
H. Wang,$^{1}$
S. R. Cui,$^{1}$
\\
$^{1}$Department of Physics and Electronic Engineering,
Yuncheng University, Yuncheng, Shanxi, 044000, People's Republic of China \\
$^{2}$Xinjiang Astronomical Observatory, Chinese Academy of Sciences,
150, Science-1 Street, Urumqi, Xinjiang, 830011, People's Republic of China \\
$^{3}$Key laboratory of Radio Astronomy, Chinese Academy of Sciences,
Nanjing, 210008, People’s Republic of China \\
$^{4}$Key Laboratory of Microwave Technology,
Urumqi, Xinjiang, 830011, People’s Republic of China \\
$^{5}$Guizhou Provincial Key Laboratory of Radio Astronomy and Data Processing,
Guiyang, Guizhou, 550001,People’s Republic of China \\
$^{6}$School of Physics and Electronic Engineering, 
Guangzhou University, 510006, Guangzhou, People's Republic of China \\
$^{7}$School of Physical Science and Technology, 
Xinjiang University, urumqi, Xinjiang, 830046, People's Republic of China \\
}

\date{Accepted 2022 December 03. Received 2022 November 21; in original form
2022 August 13}

\pubyear{2022}

\begin{document}
\label{firstpage}
\pagerange{\pageref{firstpage}--\pageref{lastpage}}
\maketitle

\begin{abstract}
In this study, we report on a detailed single pulse analysis of the radio
emission from the pulsar J1401$-$6357 (B1358$-$63) based on data
observed with the ultrawideband low-frequency receiver on the Parkes radio
telescope.
In addition to a weak leading component, the integrated pulse profile 
features a single-humped structure with a slight asymmetry.
The frequency evolution of the pulse profile is studied.
Well-defined nulls, with an estimated nulling fraction greater than 2\%, 
are present across the whole frequency band.
No emission is detected with significance above 3$\sigma$ in the
average pulse profile integrated over all null pulses.
Using fluctuation spectral analysis, we reveal the existence of
temporal-dependent subpulse drifting in this pulsar for the first time.
A clear double-peaked feature is present at exactly the alias border
across the whole frequency band, which suggests that the apparent drift sense
changes during the observation.
Our observations provide further confirmation that the phenomena of pulse
nulling and subpulse drifting are independent of observing frequency, which
suggest that they invoke changes on the global magnetospheric scale.
\end{abstract}

\begin{keywords}
	stars: neutron -- pulsars: -- general -- pulsars: individual (J1401$-$6357)
\end{keywords}

\section{Introduction}
\label{sec:int}

Pulsars are famous as precise clocks in the areas of fundamental physics under
extreme conditions \citep{Deng+etal+2020,Deng+etal+2021,Gao+etal+2021,Yan+etal+2021},
which are attributed to the exceptionally regular rotation and stable integrated pulse
profiles.
The integrated pulse profiles derived from averaging over tens of thousands
of pulses are highly stable for the majority of pulsars, and remain
stable temporally.
In contrast, a myriad of intrinsic amplitude and phase modulation effects are
displayed in the sequential rotations of a pulsar, which directly reflects the 
variation in the magnetosphere.
Occasionally, the pulse emission quenches abruptly for a period of time, which 
has been dubbed as nulling \citep{Backer+1970a}, and the phenomenon has
been detected in more than 200 pulsars.
Pulse nulling appears to occur at all frequencies and to all components
simultaneously, but partial nulls are also detected \citep{Wang+etal+2007}.
A variety of models have been proposed for nulling, such as orbital companions
\citep{Cordes+Shannon+2008}, missing line of sight
\citep{Herfindal+Rankin+2007}, switching between curvature radiation and inverse 
Compton scattering \citep{Zhang+etal+1997}, non-radial oscillations
\citep{Rosen+etal+2011}, pressional torques \citep{Jones+2012}, magnetic field
instability \citep{Geppert+etal+2003}, switching between magnetospheric
states \citep{Timokhin+2010}, and the modification of surface magnetic field
\citep{Geppert+etal+2021}.
Amongst the most promising theories is the possibility that the pulsar
magnetosphere experiences systematic variation in the global charge
distribution \citep{Lyne+etal+2010}.
In some pulsars, the subpulses were observed to perform systematic
drift motion within the pulse window, which is termed as subpulse drifting
phenomenon \citep{Drake+Craft+1968}.
The periodic phase-modulated drifting appears to be a universal behaviour, and
has been reported in at least one-third of the normal pulsars
\citep{Weltevrede+etal+2006} and in some millisecond pulsars
\citep{Edwards+Stappers+2003}.
The sparking model for subpulse drifting as suggested by
\citet{Ruderman+Sutherland+1975} has been widely accepted, in which a
number of sparks discharge in a gap over the polar cap and circulate around the
magnetic pole in the pulsar magnetosphere.
Recently, several studies have highlighted the limitations for the
interpretations of sparking discharges.
\citet{van+Timokhin+2012} suggested that changes in the drift velocity are
dependent on the variation in the accelerating potential across the polar cap,
and employed multipoles to model the subpulse drift reversals.
\citet{Wright+Weltevrede+2017} explained the phenomenon of bi-drifting subpulses
by suggesting that the carousel beams are elliptical with axes tilted 
relative to the fiducial plane.
\citet{Szary+van+2017} introduced the effect of non-dipolar magnetic fields, and
proposed that the plasma in the inner acceleration region rotates around the
point of maximum potential at the polar cap.
\citet{Basu+etal+2020b,Basu+etal+2022} suggested that the variable drift of the
sparks in the partially screened gap lags behind the corotation motion of the
pulsar.
\citet{Wright+2022} proposed that the emission patterns are originated from an
internal system with beat frequency between the magnetospheric drift and the
delay time in the rotation of the magnetosphere.
The pulse nulling and subpulse drifting phenomena are important in shaping
ideas concerning the pulsar radio emission process.

PSR J1401$-$6357 is a normal pulsar with rotational period of $P_1=0.84$ s.
It has a narrow pulse profile with a duty cycle of 4.72\% \citep{Hobbs+etal+2004}.
The linear polarization has been shown to be moderate with a gentle variation in
position angle \citep{Qiao+etal+1995}.
On the contrary, the more sensitive polarization behaviour reported by
\citet{Johnston+Kerr+2018} shows large variations in the polarization position
angle, indicating a core single proﬁle.
Infrequent short nulls are observed, which give a nulling fraction (NF) of
just 1.6\% at 1369 MHz \citep{Wang+etal+2007}.

In this paper, we focus on the properties of the individual pulse emission
from PSR J1401$-$6357 across ultrawide frequency band.
In Section~\ref{sec:obs}, we describe the observations, which used the ultrawide
bandwidth low-frequency (UWL) receiver on the Parkes radio telescope, the 
data reduction, and the data analysis.
Investigation of the average pulse profiles, pulse nulling and the
subpulse drifting is presented in Section~\ref{sec:res}.
Discussion and conclusions are given in Sections~\ref{sec:dis} and \ref{sec:con},
respectively.

\section{Observations and data processing}
\label{sec:obs}
The analyses in this paper are based on available observations from the Parkes
pulsar data archive\footnote{\url{https://data.csiro.au}} \citep{Hobbs+etal+2011}, 
which were carried out using the UWL receiver on the Parkes 64 m radio telescope on
26 November 2018.
The frequency-dependent system-equivalent flux density ranges from 33 to 72 Jy
\citep{Hobbs+etal+2020}.
The UWL radio frequency band (704$-$4032 MHz) is subdivided into 3328$\times$1 MHz
frequency channels.
The four polarization-product data are processed by the `Medusa' graphics processing
unit cluster and written to disc in 
PSRFITS\footnote{\url{http://www.atnf.csiro.au/research/pulsar/psrfits/}} search-mode
format \citep{Hotan+etal+2004} with 8-bit quantization.
The pulsar was successively observed with time resolution of 64 $\mu$s for around 1 hour.

In the off-line data processing, the raw filterbank data were initially processed to 
mitigate spurious radio frequency interference (RFI) signals.
The frequency channels at subband boundaries were zero-weighted.
The RFI-affected frequency channels were identified and flagged using a
median-filter technique.
Subsequently, the data from each channel were incoherently dedispersed at the
dispersion measure of the pulsar (DM=98.0\pcm) to correct for the
frequency-dependent spread caused by the dispersive effects of the interstellar plasma.
After that, the data were averaged into eight separated subbands, each
having an equal bandwidth of 416 MHz, to study the frequency evolution of the
pulsar emission.
Finally, using the pulsar ephemerides from the ATNF pulsar 
catalogue\footnote{\url{http://www.atnf.csiro.au/people/pulsar/psrcat/}}
\citep{Manchester+etal+2005}, the data were re-sampled and folded at the topocentric 
spin period to extract a three-dimensional pulse-stack with 1024 longitude bins on 
the x-axis, 4210 pulses on the y-axis, and 8 frequency subbands on the z-axis.

\begin{figure*}
	\centering
	\includegraphics[width=18.0cm,height=20.0cm,angle=0,scale=1.0]{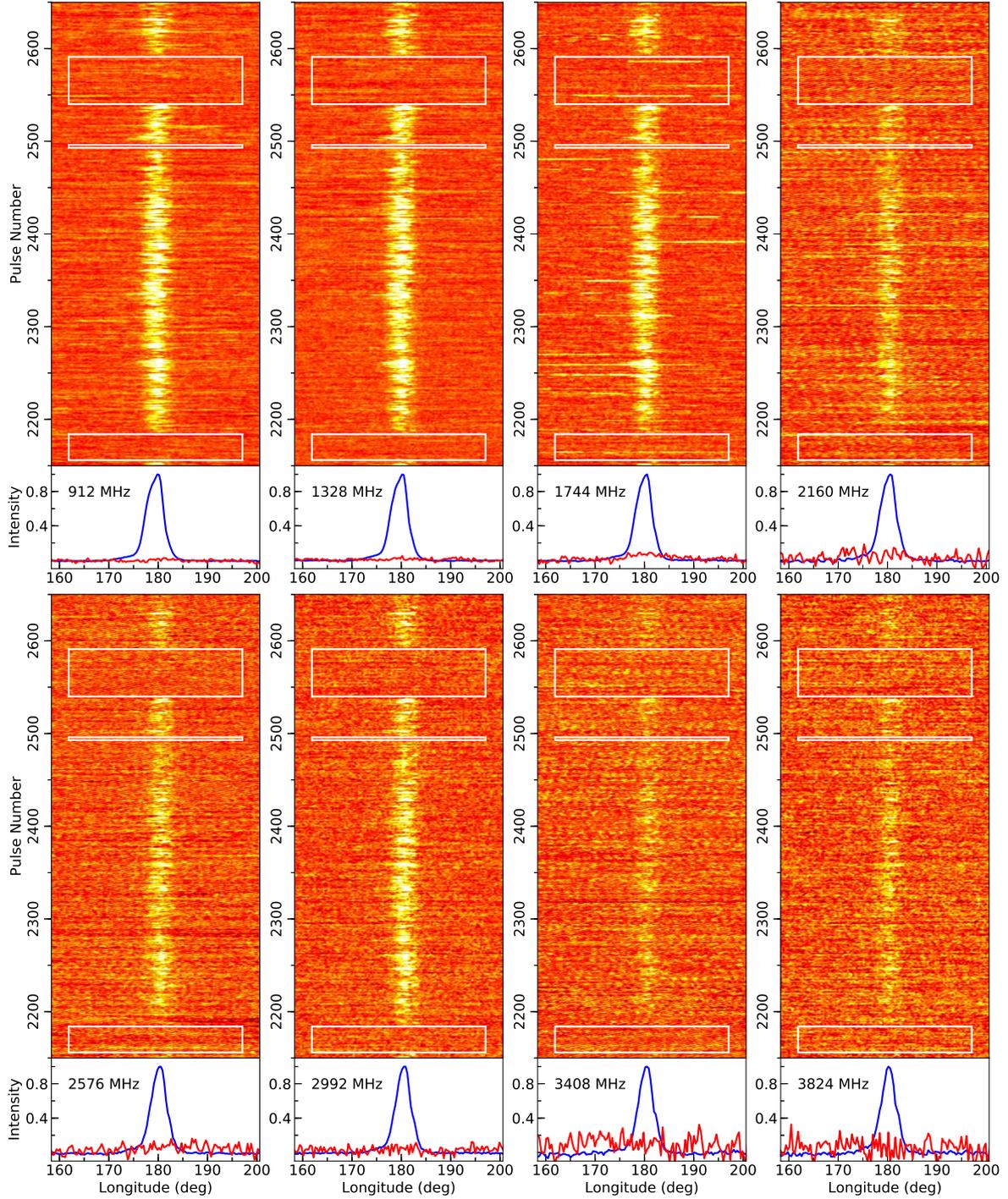}
	\caption{Plots showing the single pulse stacks for PSR J1401$-$6357 at
	eight different frequencies across the UWL band, with their corresponding
	average pulse profiles (blue) shown in the bottom panels.
	The three intervals of quenching emission are indicated by the white
	rectangles in each plot.
	The averaged null pulse profiles are shown with red lines.
	No significant emission is present in the null pulse profiles.
	The average profiles are normalized by the peak amplitude of the burst
	profiles.}
	\label{pic:sgl}
\end{figure*}

As an example of the output of the data processing, the single pulse sequences
and their corresponding average pulse profiles at eight different frequencies 
across the band of the UWL are shown in Figure~\ref{pic:sgl}.
Prominent burst pulses are interspersed with nulls at all frequencies concurrently.

\section{Results}
\label{sec:res}

\subsection{Average pulse profile}
\begin{figure}
	\centering
	\includegraphics[width=8.0cm,height=10.0cm,angle=0,scale=1.0]{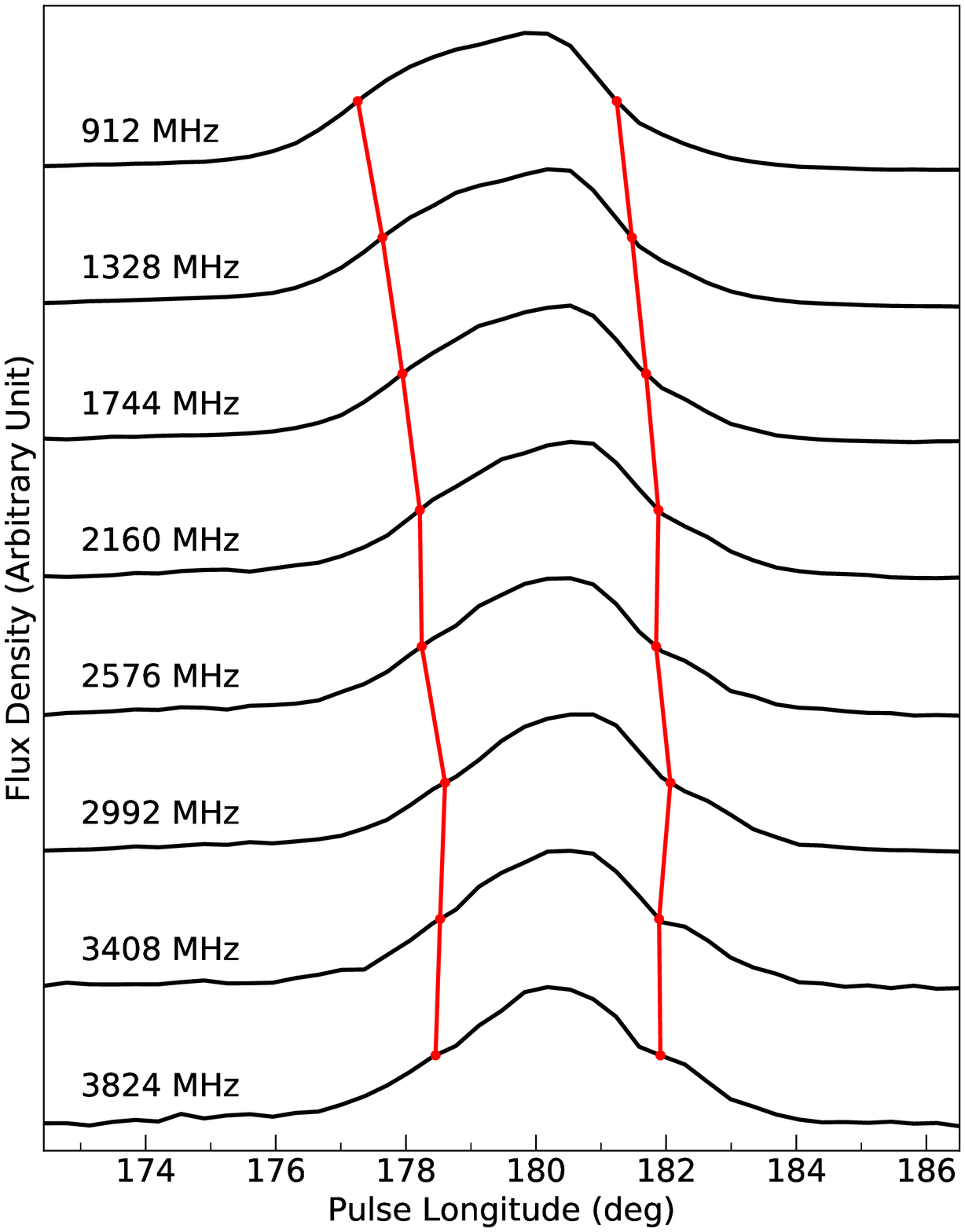}
	\caption{Broadband pulse profiles of PSR J1401$-$6357.
	The black solid lines represent the observed profiles.
	The red dots indicate the $W_{50}$ of the pulse profiles.}
	\label{pic:ps_line}
\end{figure}

The behaviour of a radio pulse profile as frequency evolves sheds insight into
the emission geometry of the pulsar, including the height of the
emission region, the beam shape, radius-to-frequency mapping (RFM) and so on.
The pulsar radiation models can be further constrained by studying the
morphologies of pulse profiles in multibands.
PSR J1401$-$6357 was reported to have a narrow pulse profile with a weak
extension in the leading side \citep{Hobbs+etal+2004}.
The prominent component is composed of a conflated core and trailing conal
outrider at 1.4 GHz \citep{Johnston+Kerr+2018}.
PSR J1401$-$6357 has been regarded as having a core-single (S$_t$)
profile \citep{Rankin+2022}, according to the morphological classification for
pulse profiles proposed by \citet{Rankin+1983}.
The average pulse profiles of PSR J1401$-$6357 at eight observing
frequencies are displayed in Figure~\ref{pic:ps_line}, where the pulse peaks
are normalized to unity.
It is remarkable that the pulse profile changes with frequency in a
characteristic manner.
As the frequency decreases, the core component becomes more and more
asymmetric in shape.
In general, the core components have widely asymmetric shapes and bifurcations.
Such effects were also observed in PSR B1933+16, which is a
core-emission-dominated pulsar, and exhibits a very clear non-Gaussian
structure with notches \citep{Mitra+etal+2016}.
At high frequencies, its two outriders become more and more intensive when
comparing with the main component \citep{Sieber+etal+1975}.
The core emission from the mode changing pulsar B0329+54 is found to be
intensity-dependent.
The stronger core emission that appears at the leading part of the component
is interpreted as due to the aberration/retardation effect \citep{Brinkman+etal+2019}.
The asymmetrical structures shown in the profile components are proposed to
originate from intrinsic effects, such as the plasma generation process and
emission mechanism, as well as propagation effects in the pulsar magnetosphere.
As a result the Gaussian fitting analysis described by \citet{Kramer+1994} often
leads to improper identification of the individual components, which is not
applicable in the present case.

\begin{figure}
	\centering
	\includegraphics[width=8.0cm,height=6.0cm,angle=0,scale=1.0]{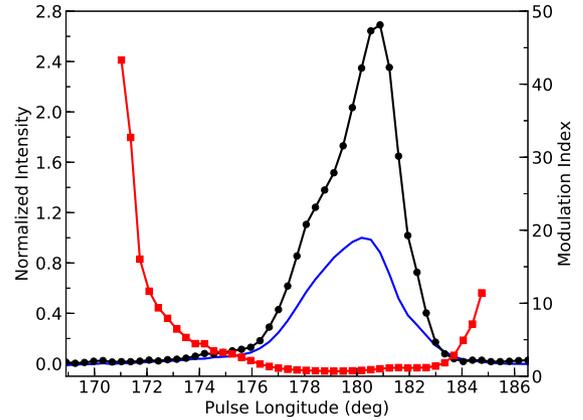} 
	\caption{The integrated pulse profile, the longitude-resolved standard
	deviation and the longitude-resolved modulation index are displayed in blue,
	black and red solid lines, respectively.}
	\label{pic:modulation_index}
\end{figure}

\begin{figure}
	\centering
	\includegraphics[width=8.0cm,height=6.0cm,angle=0,scale=1.0]{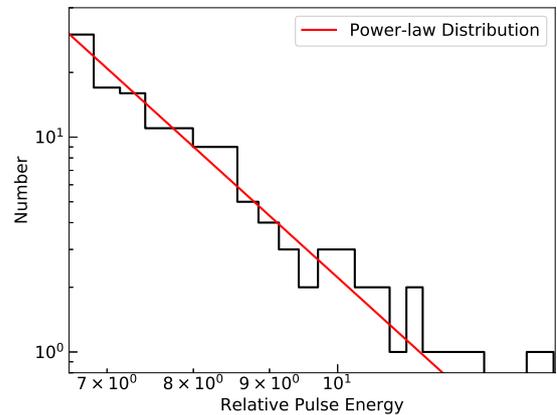} 
	\caption{Relative pulse energy distribution for the leading edge of the profile.
	To obtain a more accurate estimate, only the pulses where the bursts
	were present are included in the distribution.
	The best-fitted power-law distribution with estimated index of $-6.3\pm$0.3
	is shown in the red line.}
	\label{pic:leading_energy}
\end{figure}

\begin{figure}
	\centering
	\includegraphics[width=8.0cm,height=6.0cm,angle=0,scale=1.0]{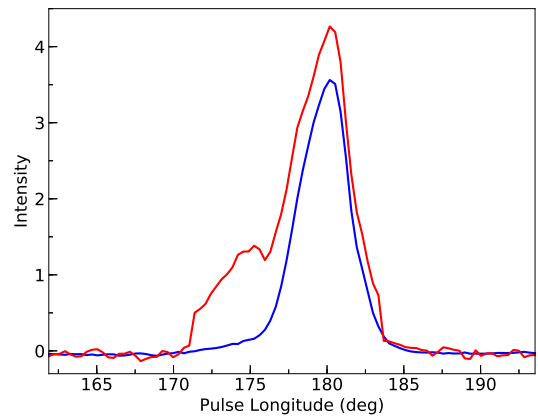} 
	\caption{The average pulse profiles in the presence (red) and absence
	(blue) of the sporadic leading part.}
	\label{pic:prof_comp}
\end{figure}

The weak leading component is located in pulse longitude ranging from
170$^\circ$ to 176$^\circ$.
By studying the individual pulse emission at 1369 MHz,
\citet{Wang+etal+2007} argued that the weak leading component was formed from
highly sporadic pulses.
In addition, the occasional pulses are comparable in flux density to those
at the main peak.
The modulation index is a measure of the factor by which the intensity
varies from pulse to pulse \citep{Weltevrede+etal+2006}.
Therefore, it is taken as an indication for the presence of subpulses.
To identify the sporadic leading outbursts, the longitude-resolved
modulation index is calculated, which is defined as $m_i=\sigma_i/\mu_i$, where
$\sigma_i$ is the standard deviation and $\mu_i$ is the mean intensity in the
$i$-th bin.
In Figure~\ref{pic:modulation_index}, the black line is the longitude-resolved
standard deviation and the red line corresponds to the longitude-resolved
modulation index.
The longitude-resolved modulation index varies a lot with pulse longitude.
It shows a minimum in the middle of the pulse profile where the total intensity
is relative high.
A sharp peak at the leading edge of the pulse profile is presented in the
longitude-resolved modulation index, which is caused by bright subpulses.
To separate pulses with the presence of sporadic leading bursts, the
relative pulse energies are calculated for the leading edge of the profile.
The occurrence of the sporadic emission in the leading part of the profile is
rather rare, with only 184 pulses out of 4210 pulses having pulse energies six
times greater than the average.
Figure~\ref{pic:leading_energy} presents the relative pulse energy distribution
for the leading edge of the profile.
To obtain a more accurate estimate, only the pulses where the bursts were
present are included in the distribution.
The best power-law fit to the pulse energy distribution gives an index of $-6.3\pm0.3$.
Figure~\ref{pic:prof_comp} shows the average pulse profiles in the presence and
absence of the sporadic leading part.
It is obvious from the figure that the profiles are different.
For the integrated pulse profile with the presence of the sporadic leading part,
the relative intensity at the leading edge of the profile is larger and the
pulse width is significantly broader.
In addition, the central component is stronger during the burst state
compared to the normal state.

\begin{figure}
	\centering
	\includegraphics[width=8.0cm,height=6.0cm,angle=0,scale=1.0]{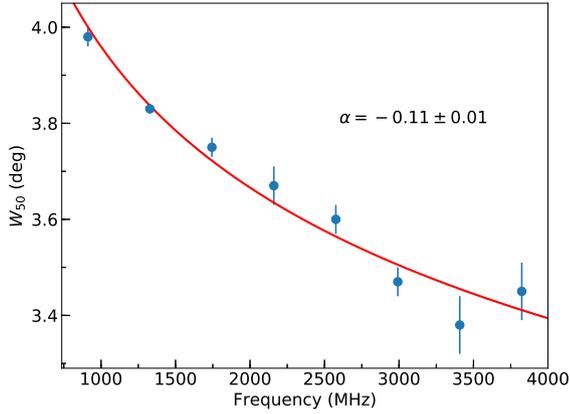}
	\caption{Evolution of the profile width with the observing
	frequency.
	The red curve represents the best power-law fit with an index estimated
	to be $\alpha=-0.11\pm0.01$.}
	\label{pic:width}
\end{figure}

\begin{figure}
	\centering
	\includegraphics[width=8.0cm,height=6.0cm,angle=0,scale=1.0]{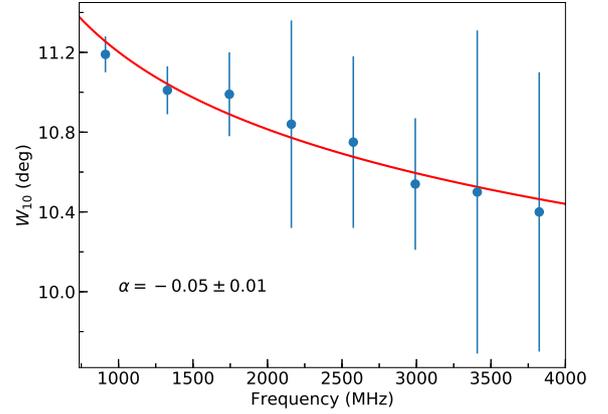}
	\caption{Frequency evolution of the profile width during the leading
	burst state.
	The red curve represents the best power-law fit with an index estimated
	to be $\alpha=-0.05\pm0.01$.}
	\label{pic:width10}
\end{figure}

The mean pulse profile width at 50\% ($W_{50}$) of the peak amplitude 
is calculated for the eight frequency bands.
Considering just one component above the 50\% level is shown in the mean pulse
profiles, the uncertainty in $W_{50}$ is estimated to be
\begin{equation}
	\centering
	\sigma_{W_{50}} = \frac{W_{50}}{\sqrt{2}\rm{ln(2)S/N}},
\end{equation}
where S/N is the signal-to-noise ratio of the profile.
Figure~\ref{pic:width} shows the frequency evolution of the profile width.
It is apparent from the figure that the value of $W_{50}$ decreases with
increasing frequency, which roughly agrees with a power-law function.
The power-law index is estimated to be $-0.11\pm0.01$.
In Figure~\ref{pic:ps_line}, the $W_{50}$ for each frequency is also
plotted.
The frequency evolution of the profile width during the leading burst state
is shown in Figure~\ref{pic:width10}.
The pulse width at 10 per cent ($W_{10}$) level of the peak intensity is
calculated since the leading part emission is below the 50 per cent level of the
peak.
The best power-law fit gives an index of $\alpha=-0.05\pm0.01$.
The pulse width measurements are available over a broad range of radio
frequencies.
The relation between pulse width and frequency for 150 normal pulsars has been studied
by \citet{Chen+Wang+2014}.
Clear pulse width narrowing with increasing frequency for about 54\%
of their pulsars were reported.
The best-fit Thorsett indices are distributed within $-1$ and 0, and peaked near 0.
\citet{Posselt+etal+2021} presented measurements of a homogeneous large sample
of pulse widths from the Thousand-Pulsar-Array programme on the MeerKAT
telescope. 
The pulse width as a function of the rotation period is described with a steep
power-law.
The power-law index is estimated to be $-0.63\pm0.06$ using orthogonal distance
regression technique.
In addition, a monotonic behaviour with frequency is shown in the measured width changes.
The core and conal component widths versus period in normal pulsars were
measured to have a lower boundary line that closely follows the power-law
relation with index of $-0.5$ \citep{Skrzypczak+etal+2018}.
\citet{Xu+etal+2021} systematically investigated the frequency evolution
behaviour of profile widths for 74 pulsars.
The power-law indices of most pulsars are measured to be around $-0.15$.
The profile evolution with observing frequency is consistent with the RFM effect, 
where the high-frequency emission is assumed to be generated at a lower altitude 
and vice versa \citep{Cordes+1978}.
The magnetospheric electron density is expected to decrease with increasing
altitude, predicting that the radio beam expands with decreasing frequency
\citep{Ruderman+Sutherland+1975}.

Measuring the altitudes of radiation locations is of great importance for the 
research of pulsar radiation mechanism as well as research for physical 
processes in pulsar magnetospheres.
The emission altitudes can be estimated by considering the geometrical
properties of the radiation beam, which is proposed by \citet{Cordes+1978} and
developed further by \citet{Kijak+Gil+1997}.
The inclination angle and impact angle are basic parameters for measuring
the altitude.
They can be obtained by fitting the linear polarization position angle curve
in the rotating vector model \citep{Radhakrishnan+Cooke+1969}.
Furthermore, the polarization behaviour serves the purpose of revealing
whether the leading part can be classified as a conal component or merely an
extension of the core component.
However, the emission geometry cannot be determined for PSR
J1401$-$6357 due to the lack of calibrated observations for polarization.

\subsection{Pulse nulling}
As shown in Figure~\ref{pic:sgl}, the pulsed emission switches simultaneously
from burst state to null state (and vice versa) across the different frequencies.
This behaviour is broadly confirmed by a visual inspection of the entire data,
which implies that pulse nulling in this pulsar appears to be broadband.

\begin{figure}
	\centering
	\includegraphics[width=8.0cm,height=6.0cm,angle=0,scale=1.0]{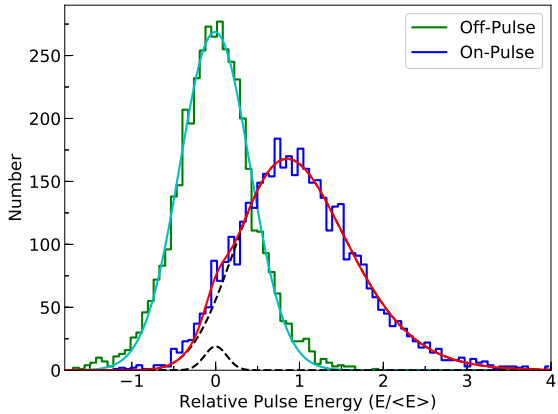}
	\caption{Pulse energy histograms for the on-pulse (blue) and off-pulse
	(green) windows.
	The energies are normalized by the mean on-pulse energy.
	The cyan line represents the off-pulse noise model fit.
	The red line indicates the distribution expected from a combination of a
	Gaussian and a lognormal distributions (black dashed lines).}
	\label{pic:energy}
\end{figure}

The investigation of pulse energy distribution provides insights into the radio
pulsar emission mechanism and the physical state of pulsar magnetospheres.
PSR J1401$-$6357 was reported to exhibit atypically broad pulse energy distribution 
at 1352 MHz \citep{Burke-Spolaor+etal+2012}.
Figure~\ref{pic:energy} presents the energy distributions for the on-pulse
and off off-pulse windows after averaged over the whole frequency band, which are
calculated using the standard method described by \citet{Ritchings+1976}.
The on-pulse window, ranging from 170$^\circ$ to 190$^\circ$, is determined
by inspection of the integrated pulse profile.
The off-pulse window is determined in the off-pulse region with the size equal 
to the on-pulse window. 
The normalized on-pulse and off-pulse energies are calculated for each
pulsar rotation by integrating the energies in the on-pulse and off-pulse
windows, respectively, and then divided by the mean on-pulse energy obtained
from all integrations.
The off-pulse energy represents the Gaussian random noise contributed by the 
system temperature, which can be well modeled with a Gaussian distribution
centered around zero.
The distribution width corresponds to the rms fluctuations of the data.
It is noted that the null and burst pulse distributions are
indistinguishable in the on-pulse energy distribution due to the low S/N burst
pulses mixing with the null pulse distribution and a small number of null pulses.
The statistical estimation on the NF is not applicable.
\citet{Burke-Spolaor+etal+2012} found that the normal pulsar population is seen
to exhibit mostly lognormal pulse energy distributions.
It appears that PSR J1401$-$6357 follows the same trend.
The Kolmogorov-Smirnov (KS) hypothesis test is applied to check whether or not 
the observed distribution can be described by a lognormal distribution.
The obtained $p$-value of the lognormality hypothesis test is greater than the
threshold value of 0.05, which indicates that the observed distribution is
consistent with a lognormal distribution. 
The lognormal statistics are intrinsic to pulsar emission, which can be
interpreted in terms of a pure stochastic growth theory (SGT)
\citep{Cairns+etal+2003}.
The associated pulsar emission mechanism is suggested to involve with only
linear processes.

\begin{figure}
	\centering
	\includegraphics[width=8.0cm,height=6.0cm,angle=0,scale=1.0]{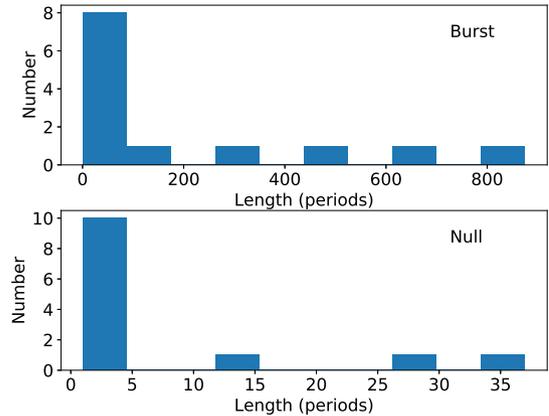}
	\caption{Histograms of observed null length and burst length for PSR
	J1401$-$6357.}
	\label{pic:duration}
\end{figure}

\begin{figure}
	\centering
	\includegraphics[width=8.0cm,height=6.0cm,angle=0,scale=1.0]{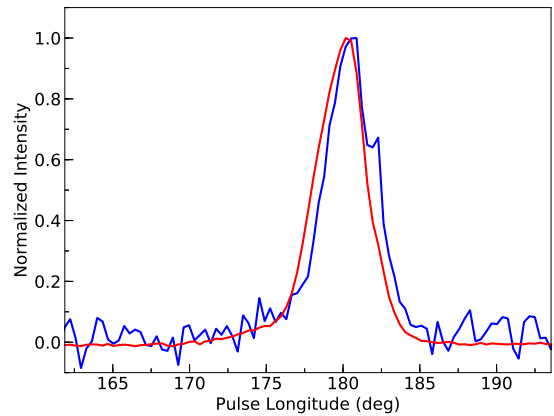}
	\caption{The average pulse profile by integrating pulses in short
	duration bursting states within the null intervals is shown in blue.
	The average pulse profile by integrating pulses in the long burst states
	is shown in red.}
	\label{pic:prof_ab}
\end{figure}

To improve the detection sensitivity for the single pulses for nulling
analysis, the signals from the different subbands are combined.
We follow the procedure presented by \citet{Bhattacharyya+etal+2010} for
identification of individual nulls, in which pulses with intensity smaller than
3$\sigma_{ep}$ are classified as nulls, where $\sigma_{ep}$ is the uncertainty
in the pulse energy calculated from the root-mean-square energy in the off-pulse
window.
To further identify the null pulses with more accuracy, a visual inspection is
carried out on the pulse sequence and the mislabeled nulls and bursts are
redressed.
There are 95 null pulses identified and the NF is estimated to be 2.26$\pm$0.23\%.
The uncertainty of NF is simply given by $\sqrt{n_p/N}$, where $n_p$ is the
number of null pulses and $N$ is the total number of observed pulses.
Our determined NF is larger than that estimated by \citet{Wang+etal+2007} at 1369 MHz.
A total of 13 blocks of burst state are identified from the whole set of
observations.
The histograms for null length and burst length are formed and displayed in
Figure~\ref{pic:duration}.
The length distribution of contiguous nulls is apparently dominated by 1-period
nulls.
The short nulls are predominantly weak normal pulses, which occur randomly among the 
emission bursts.
There are three long null states with durations of 13, 28 and 37 pulses.
The durations of burst states vary from 1 pulse to 1514 pulses.
The occasional individual strong pulses are found within the null intervals.
Figure~\ref{pic:prof_ab} shows the normalized pulse profiles obtained by
integrating pulses in short (blue) and long (red) bursting states.
The profile of short duration bursts is wider than that of long duration bursts.
In comparison with the profile for the long burst state, the profile for short
duration bursts shifts to later longitude by almost 0.7$^\circ$.

\subsection{Subpulse drifting}

\begin{figure}
	\centering
	\includegraphics[width=8.0cm,height=10.0cm,angle=0,scale=1.0]{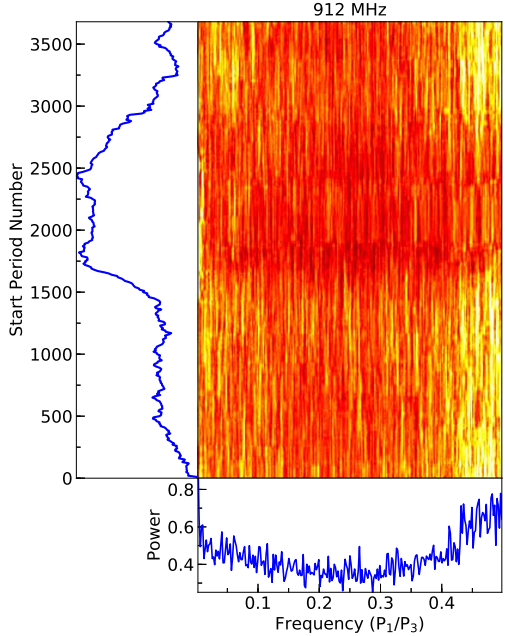}
	\caption{Main: The time-varying longitude-resolved fluctuation spectra.
	Bottom: The averaged spectrum across all times showing the presence of
	high-frequency modulation feature.
	Left: The time variation of the modulation power.}
	\label{pic:lrfs_tim}
\end{figure}

In general, drifting subpulses, nulling, and mode changing can be detected 
through visual inspection of the single pulse sequence for bright pulsars.
\citet{Backer+1973} and \citet{Backer+etal+1975} developed a systematic approach to 
measure the drifting features, including the fluctuation spectral studies.
This often-used technique is suitable for detecting subpulse drifting in weak
pulsars.
The drifting is characterized by two periodicities: $P_2$ (the longitudinal separation
between two adjacent drift bands) and $P_3$ (the interval over which the subpulses
repeat at a specific location within the pulse window).
To detect a regular intensity variation, the longitude-resolved fluctuation
spectrum (LRFS) is calculated to estimate $P_3$ \citep{Backer+1970b}.
To do that, the pulse-stack is divided into blocks of 512 successive
pulses, and then the discrete Fourier transform (DFT) is performed along each
longitude bin within the pulse window on these blocks.
The final spectrum is obtained by averaging the fluctuation power across all
longitudes along the pulse window.
An example is presented in Figure~\ref{pic:lrfs_tim}, which is used 
to determine the time evolution of periodic modulation at 912 MHz.
The drift tracks with regions of enhanced intensity are visible, which 
indicate the changes in periodic modulation with time.
The lower panel shows the averaged power spectrum across all blocks, which 
also demonstrates the presence of a high-frequency modulation
feature ranging from 0.4 to 0.5 cycle period$^{-1}$.
The time variation of the modulation power is shown in the left panel, where the 
periodic modulation vanishes between pulse numbers 1500 and 2750.
The temporal subpulse drifting behaviour across multiple frequencies is
investigated in order to reveal the simultaneity in the
changes of the subpulse drifting across the different frequencies,
which will also reveal the broadband nature of the subpulse drifting in this pulsar.

\begin{figure}
	\centering
	\includegraphics[width=8.0cm,height=12.0cm,angle=0,scale=1.0]{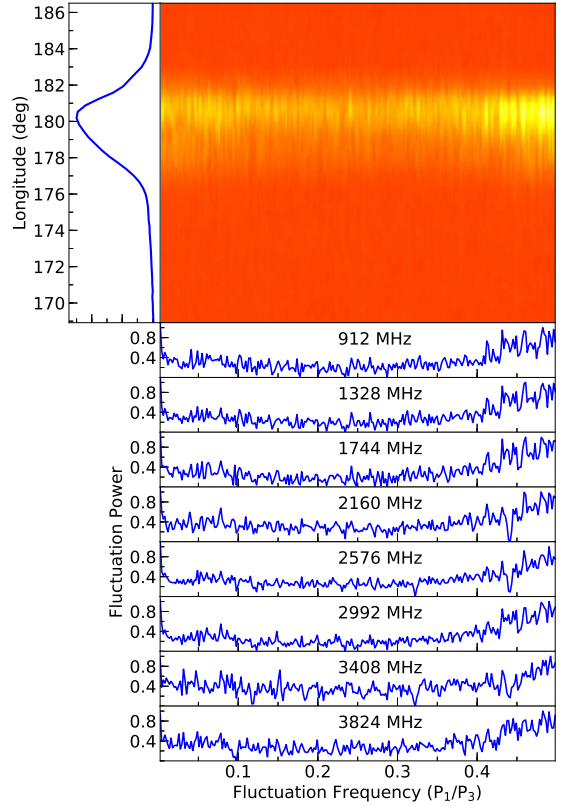}
	\caption{Main: The longitude-resolved fluctuation power spectra averaged over
	the whole observing frequency band.
	The abscissa is the frequency of the drifting periodicity and the ordinate is
	the profile longitude.
	The intensity of the drifting periodicities as a function of pulse longitude 
	is plotted in color contours.
	The total power profile is shown in the left panel.
	The integral power spectra at the eight frequencies are presented in the lower
	panel.}
	\label{pic:lrfs}
\end{figure}

To describe drifting in an averaged sense, the LRFS averaged over the whole 
observing frequency band is shown in the main panel of Figure~\ref{pic:lrfs}.
A distinct region of power excess is identified in the corresponding pulse
longitude from 178$^\circ$ to 182$^\circ$.
The high-frequency structured subpulse modulation feature appears to range from
0.4 to 0.5 cycle period$^{-1}$ ($P_1/P_3$) at the eight observing frequencies.
The pulsar is classified as a diffuse drifter, because the broad modulation 
feature is not separated from the alias border clearly \citep{Weltevrede+etal+2006}.

\begin{figure}
	\centering
	\includegraphics[width=8.0cm,height=12.0cm,angle=0,scale=1.0]{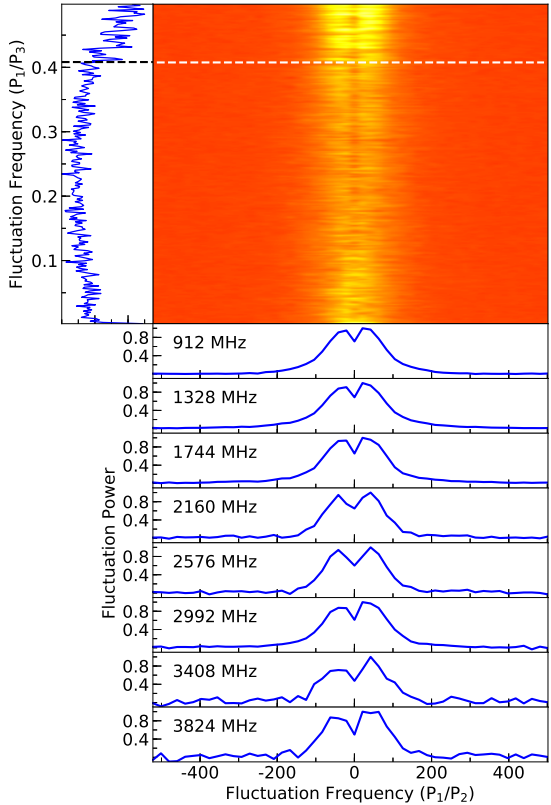}
	\caption{Main: Two-dimensional fluctuation power spectra averaged over the
	whole observing frequency band.
	The units of the horizontal axis are $P_1/P_2$, where $P_2$ is the
	horizontal separation between drift bands in time units.
	The power in the 2DFS is horizontally integrated, producing the left power
	spectra.
	The vertically integrated power spectra at the eight frequencies are
	presented in the lower panel.}
	\label{pic:2dfs}
\end{figure}

To determine whether the subpulses are amplitude-modulated drifting or
phase-modulated drifting, the two-dimensional fluctuation spectrum (2DFS) is
calculated by performing two-dimensional DFT along the vertical lines and
the lines with various slopes \citep{Edwards+Stappers+2002}.
Following the same procedures for calculating the LRFS, the pulse-stack is 
divided in blocks of 512 successive pulses and the spectra of the different blocks
are then averaged to obtain the final spectra.
In Figure~\ref{pic:2dfs}, the 2DFS averaged over the whole observing frequency
band is shown in the main panel.
The left panel shows the horizontally integrated power. 
The main feature at the same horizontal position as in the LRFS is presented, 
corresponding to the same $P_3$ value.
To determine the value of $P_3$, a similar method described by
\citet{Weltevrede+etal+2006} is adopted, which involves calculating the
centroid of a rectangular region in the 2DFS containing the feature.
The uncertainty is estimated from the power in a region, which
contains only noise, in the 2DFS.
The corresponding drifting periodicity is estimated to be $P_3 = 2.3\pm0.3 P_1$
at the eight frequencies.
The pattern repetition frequency along the pulse longitude is denoted in the horizontal
axis of the 2DFS, which is expressed as $P_1/P_2$.
It is noted that a clear double-peaked feature in 2DFS is present, which
indicates that the apparent drift sense changes during the observation.
The values of $P_2$ are estimated to be $-7\pm3$ and $8\pm3$ degrees, which
exceed the pulse width.
This means that only one subpulse is visible in a single pulse, and the drifting
appears as an amplitude modulation rather than as a phase
modulation \citep{Asgekar+Deshpande+2005}.
The drift bands often appear very nonlinear, and therefore the magnitude
of $P_2$ probably possesses little meaning.
Alternatively, the subpulse phase jump or swing could produce the double peaked
feature \citep{Edwards+etal+2003}.
The fluctuation spectra at the eight different frequencies are the
same, which imply that the subpulse drifting is independent of the
observing frequency.

In order to rule out the possibility that the occasional occurrence of strong 
subpulses and RFI are dominating the spectra and therefore lead to misleading 
conclusions, a similar method is adopted to further demonstrate the authenticity
of the periodic ﬂuctuations \citep{Weltevrede+etal+2006}.
This involves randomizing the order of the pulse sequence, and then the
LRFSs and 2DFSs are calculated from the newly formed pulse stack. 
No well-deﬁned $P_3$ in this process is detected in the emission window and 
the noise window, which confirms the signiﬁcance of the drift features.

Recently, the quasi-periodic fluctuation features are broadly categorized
into two groups: longitude-modulated drifting and periodic amplitude modulation
\citep{Basu+etal+2019a}.
Subpulse drifting is shown to be profile dependent, with the phenomenon found in
the conal components and absent in the central core emission \citep{Basu+etal+2019a}.
The amplitude-modulated drifting is thought to be a different phenomenon from
subpulse drifting \citep{Basu+etal+2016,Basu+etal+2017}.
A survey of the periodic amplitude modulation and periodic nulling has been
carried out in the pulsar population by \citet{Basu+etal+2020c}.
These longitude-stationary behaviours appear to occur across the entire profile in
both the core and conal components simultaneously.
The core component in some pulsars is suggested to exhibit long periodic
fluctuation structure, which is not classified as subpulse drifting
\citep{Rankin+1986}.
Pulsars with core-single profiles are found to exhibit diffuse fluctuation
features with no discernible longitude shift \citep{Basu+etal+2016}.
Highly periodic longitude-stationary amplitude modulation is revealed to be
temporal-dependent in the core-single radio pulsar B1946+35 \citep{Mitra+Rankin+2017}.
The longitude-stationary non-drift amplitude modulation in PSR J1048$-$5832 is
argued to be associated with periodic mode changing \citep{Yan+etal+2020}.
The pulsars B2000+40, J2048$-$1616 and J2006$-$0807 with core-cone profiles show
the presence of periodicity in association with nulling as well as subpulse drifting
\citep{Basu+etal+2020a,Wang+etal+2021,Basu+etal+2019b}.
The fluctuation spectral analysis reveals that the shorter modulation periodicities 
are associated with the subpulse drifting and absent in the central core component. 
The longer periodicities are associated with the periodic nulling behaviour.
Therefore, the detected periodic behaviour in the core-single pulsar
J1401$-$6357 can be interpreted as longitude-stationary amplitude modulation.

\section{Discussion}
\label{sec:dis}

In general, the observed drifting periodicity can either be the true drifting 
periodicity for complete Nyquist sampling or an aliased drift periodicity for
incomplete Nyquist sampling.
In a sparking gap model, the spark-associated plasma columns drift due to an
$\it{E \times B}$ drift, which depends on both the pulse period and its
derivative \citep{Ruderman+Sutherland+1975}.
In the presence of aliasing, it is impossible to determine the physical 
conditions of the pulsar from the apparent drift rate.
The lower value of $P_3$ can result from aliasing of a higher $\it{E \times B}$ 
drift, and vice versa.
As shown in Figure~\ref{pic:2dfs}, the feature in the 2DFS is not clearly
separated from the alias border, and the power in that feature could 
come from drift bands of different alias modes.
It is a challenge to relate the measured drifting periodicity to the true drift
rate of the emission sparks.
A visual inspection is carried out on the pulse sequence, which reveals that
the apparent drift direction is constantly changing during the observation.
\citet{Weltevrede+etal+2006} simulated a pulse sequence with drifting that
constantly crosses the $P_3=2P_1$ alias border to illustrate this effect.
There are a number of pulsars that show evidence for drift reversals.
For example, PSR B2303+30 shows a clear double peaked feature in its 2DFS
\citep{Redman+etal+2005}.
The feature follows a steady even-odd pattern in the intense mode B
emission, which is explained by the constantly changing drift patterns in terms
of varying subbeam rotation rates.
PSR B0826$-$34 is reported to show drift reversals, and it continuously changes 
the apparent drift direction via longitude stationary subpulse modulation
\citep{Gupta+etal+2004,Esamdin+etal+2005}.
Furthermore, there is no evidence for the existence of a pulsar 
sub-population with subpulse drifting that locates close to the $P_3=2P_1$
Nyquist limit \citep{Wright+2003}. 

We find that pulse nulling and subpulse drifting phenomena in PSR J1401$-$6357
are consistent across the ultrawide frequency band.
The occurrence of pulse nulling appears highly concurrent 
across the different frequency bands, which is consistent with the behaviour
reported by \citet{Gajjar+etal+2014}. 
A systematic and simultaneous multi-frequency study of pulsars shows that
subpulse drifting and pulse nulling are broadband in nature \citep{Naidu+etal+2017}.
The switching between multiple drift rates with distinct $P_3$ occurs
simultaneously across all the observed frequencies.
The broadband emission related to the localized emission processes is proposed
to occur close to the pair production fronts near the polar caps
\citep{Melrose+1996}.
Intrinsic models invoking extinction of the sparking region are proposed,
in which the entire magnetosphere is undergoing rapid changes causing
cessation of the radio emission.
There is increasing evidence that the cessation of emission is associated with the
disruption of the entire particle flow.
The switching between the radio-quiet `off' and radio-loud `on' states in the 
intermittent pulsar B1931+24 was found to be correlated with the slowing down rate 
of the pulsar \citep{Kramer+etal+2006} .
\citet{Lyne+etal+2010} presented the correlated changes between the shape
of the pulse proﬁle and the timing noise for six pulsars.
Furthermore, the implied broadband nature of pulse nulling and subpulse
drifting also suggests that the geometry of pulsar emission, including variations
in the emission height, is unlikely to affect the behaviour of the
two phenomena.
The nulling and drifting are probably governed by the gap potential and magnetic 
field in the polar gap, although counterexamples are reported in
multifrequency observations.
Different $P_3$ values at 21 cm and 92 cm for PSRs J1822$-$2256, J1901$-$0906, 
B1844$-$04, B2016+28, and B2045$-$16 are suggested
\citep{Weltevrede+etal+2006,Weltevrede+etal+2007}.
The frequency dependence of a well known drifter PSR B0031$-$07 was investigated
by \citet{Smits+etal+2005}.
In the context of the carousel model, several ideas have been advanced to
explain this frequency-dependent behaviour of subpulses.
\citet{Davies+etal+1984} suggested that the magnetic field is not purely
dipolar, but it includes a multipole component with contra-rotation
twisting at large radii.
\citet{Edwards+Stappers+2003} proposed the ﬁnite size and shape of beamlets 
resulting in their cross sections being slightly shifted relative to
the line of sight as the frequency-dependent beamlets move in magnetic colatitude.
Recently, \citet{McSweeney+etal+2019} developed an idea to obtain quantitative
predictions of how subpulses shift with frequency by speculating that aberration
and retardation effects are the dominant cause.

\section{Conclusions}
\label{sec:con}
Using the highly sensitive ultrawideband Parkes observations, we 
have performed detailed single pulse analysis for PSR J1401$-$6357.
From the ﬂuctuation spectral analysis, this pulsar is identiﬁed as a 
diffuse subpulse drifter for the first time. 
The subpulse modulation is extended toward the $P_3 = 2 P_1$ alias border, and
the 2DFS shows a feature that is split by the vertical axis owing to the 
two apparent drift directions in the pulse sequence.
Our results confirm and further strengthen the conclusions that pulse nulling
and subpulse drifting are broadband phenomena.
One hour single pulse observations are sufficient for the identification of these 
phenomena themselves.
Our ultrawideband observations of the individual pulses from PSR J1401$-$6357
contribute to the generalization of the broadband behaviour of pulse nulling
and subpulse drifting.

\section*{Acknowledgements}
We would like to thank the anonymous referee for providing constructive
suggestions, which helped in improving the paper.
This work is partially supported by the National Key Research and Development
Program of China (No. 2022YFC2205203), the National Natural
Science Foundation of China (NSFC grant No. 11988101, U1838109, 12041304,
12041301, 11873080, U1631106), the National SKA Program of China
(2020SKA0120100), the Chinese Academy of Sciences Foundation of the young
scholars of western (No. 2020-XBQNXZ-019), and Xinjiang Key Laboratory of
Radio Astrophysics (No. 2022D04058).
Z.G.W. is supported by the 2021 project Xinjiang uygur autonomous region of China for
Tianshan elites. 
J.L.C. is supported by the Natural Science Foundation of Shanxi Province 
(20210302123083), and the Scientific and Technological Innovation Programs of Higher 
Education Institutions in Shanxi (grant No. 2021L470).
H.W. is supported by the Scientiﬁc and Technological Innovation Programs of
Higher Education Institutions in Shanxi(Grant No.2021L480).
W.M.Y. is supported by the CAS Jianzhihua project.
H.G.W. is supported by the 2018 project of Xinjiang uygur autonomous region of China 
for flexibly fetching in upscale talents. 
S.R.C is supproted by the student innovation project of Yuncheng university (20220940).
The Parkes radio telescope, `Murriyang', is part of the Australia Telescope National 
Facility which is funded by the Australian Government for operation as a National 
Facility managed by CSIRO.

\section*{Data Availability}
The data underlying this article are available in the Parkes Pulsar Data Archive
at https://data.csiro.au, and can be accessed with the source name J1401$-$6357.

\bibliographystyle{mnras}
\bibliography{J1401-6357}

\bsp	
\label{lastpage}
\end{document}